# Bipolar nanochannels: The effects of an electroosmotic instability. Part II: Time-transient response

Ramadan Abu-Rjal and Yoav Green*
Department of Mechanical Engineering, Ben Gurion University of the Negev, Beer-Sheva 8410501, Israel
* Corresponding author. E-mail: yoavgreen@bgu.ac.il

Abstract

The most common method to characterize the electrical response of a nanofluidic system is through its steady-state current-voltage response. In Part I, we demonstrated that this current-voltage response depends on the geometry, the layout of the surface charge, and the effects of advection. We demonstrated that each configuration has a unique steady-state signature. Here, we will elucidate the behavior of the time-transient response. Similar to the steady-state response, we will show that each configuration has its own unique time-transient signature when subjected to electroosmotic instability. We show that bipolar systems behave differently than unipolar systems. In unipolar systems, the instability appears only at one end of the system. In contrast, in bipolar systems the instability will either appear on both sides of the nanochannel or not at all. If it does appear on both sides, the instability will eventually vanish on one or both sides of the system. In Part I, these phenomena were explained using steady-state considerations of the behavior of the fluxes. Here, we will examine the time-transient behavior to reveal the governing principles that are, on the one hand, not so different from unipolar systems and, on the other hand, remarkably different.

**Impact Statement.** Permselective nanoporous materials are ubiquitous in desalination, energy harvesting, and bio-sensing systems. Of particular importance are bipolar membranes and nanochannels that are comprised of two oppositely charged permselective regions. While a plethora of experimental works have characterized the electrical response of these systems, a fundamental understanding of the underlying physics determining the response is still missing. To address this knowledge gap, we have systematically simulated different bipolar nanofluidics systems subject to varying potential drops and characterized their electrical response to reveal signatures that are unique to every system. Our findings contribute to a more profound understanding of the various control parameters and mechanisms that determine the time transient dynamics and the steady-state current-voltage response in bipolar systems and provide a valuable tool for interpreting experimental and numerical data of such systems. The insights from this work can be used to improve the design of fabricated bipolar devices.

# 1. Introduction

The transport of ions across permselective materials is immensely important for numerous applications, including, but not limited to, desalination (Nikonenko et al., 2014; Marbach & Bocquet, 2019), energy harvesting (Siria et al., 2013; Wang et al., 2023), biomolecule sensing (Slouka, Senapati, & Chang, 2014; Vlassiouk, Kozel, & Siwy, 2009), and fluidic-based electrical circuits (Vlassiouk, Smirnov, & Siwy, 2008; Yossifon, Chang, & Chang, 2009; Lucas & Siwy, 2020; Sebastian & Green, 2023; Noy & Darling, 2023). However, the electrokinetic transport of ions depends on the geometry of the system, the distribution of the surface charge density, the electrolyte concentration, the applied voltage drop, and many other parameters, such that it is virtually impossible to completely characterize the electrical response of the system for the full parameter space. The difficulty of characterizing the system is further frustrated by the fact that the governing equations of the transport are a set of coupled, nonlinear partial differential equations that require numerical evaluation.

Naturally, this has led to the reliance on either experiments or simplified numerical simulations (that focus on very specific systems with a small parameter space). The goal of this two-part work has been to provide a robust and systematic scan of a large parameter space to demonstrate how the electrical response of a permselective system varies as one particular parameter is varied. To that end, we divided the results and analysis into two. In Part I (Abu-Rjal & Green, 2024), we focused on the much more studied and intuitive steady-state characteristics. Here, in Part II, we will focus on the time-transient result leading up to the steady-state response. Similar to Part I, here too, we will show that every system considered has a different signature.

As in Part I, we distinguish between three different systems – the unipolar system and the bipolar system where the bipolar system is then subdivided into two: the ideal and non-ideal. The difference between the unipolar and bipolar systems is the distribution of the surface charge density. **Figure 1** shows a schematic of a bipolar system comprised of two diffusion layers and two oppositely charged selective regions. When considering a unipolar system, one needs to remove one of the charged regions (without loss of generality, we will remove the positively charged surface by setting $L_3 = 0$ or $N_3 = 0$). At the two ends of the system are two electrodes under an applied voltage drop of $V$, which measure a current, $I$ (or vice versa, an applied current and measured voltage). Each of these systems is subjected to a wide range of voltages without and with the effect of electroosmotically driven advection while we characterize $I(t)$. To remove the dependency of the current, $I$, on the width, $W$, and height, $H$, of the system, we will consider the current density, $i(t)$ [discussed further below].

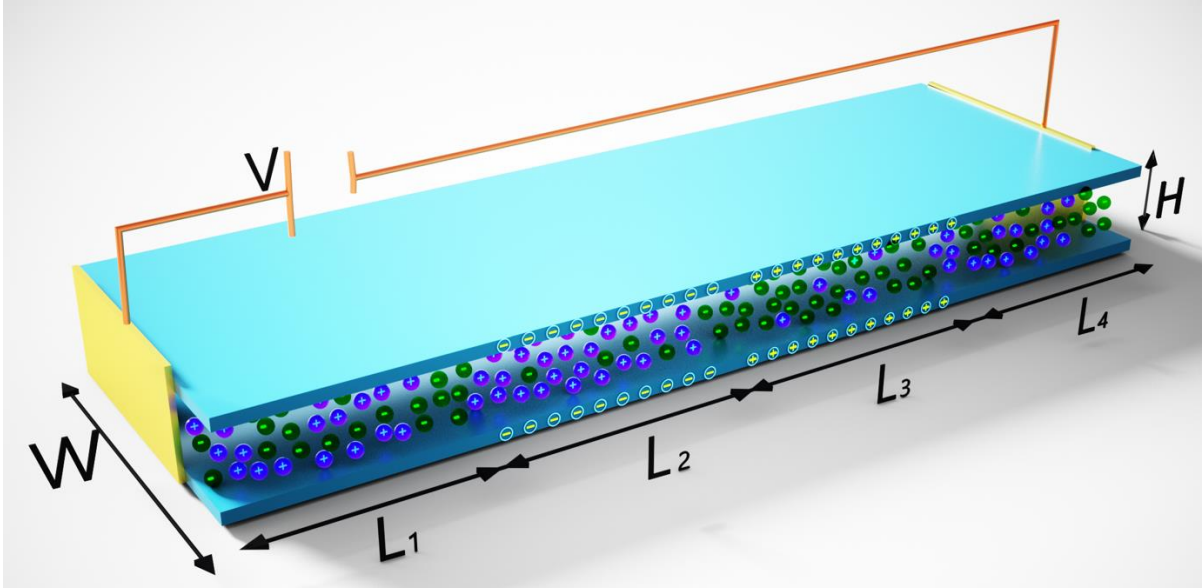

*Figure 1. Schematic of a three-dimensional four-layered system comprised of two diffusion layers connected by two permselective mediums under an applied voltage drop, $V$. The length of each of the four regions ($k = 1, 2, 3, 4$) is given by $L_k$, the height is $H$ and the width is $W$. The two outer regions are uncharged such that the concentrations of the positive ions (purple spheres) and the negative ions (green spheres) are the same. The two middle regions are charged with either a negative or positive surface charge density, leading to a surplus of counterions over coions. In the negatively charged region, the positive ions are the counterions, while in the positively charged regions, the negative ions are the counterions.*

In Part I, we demonstrated that the steady-state current density-voltage response $\langle \bar{i} \rangle - V$ of unipolar and bipolar systems differed from each other such that each configuration had a unique steady-state signature. In particular, the differences are by far more drastic when the effects of electroosmotic flows (EOF) are accounted for. Specifically, we showed that in contrast to unipolar systems, where over-limiting currents (OLCs) are always observed, in "ideal" bipolar systems (the definition for "ideal" is given below), OLCs are not observed at all. Further, any "non-ideal" bipolar system displays characteristics of both unipolar and ideal bipolar systems, where the exact response is dependent on the degree of non-ideality. The three curves are shown in **Figure 2**. In this Part, we will explain these steady-state results by considering the time-transient response of the current subjected to a constant voltage applied at $t = 0$. We will show that the appearance of the electroosmotic instability (EOI) and whether it is sustained for long times, will depend on whether at least one initial depletion layer is formed in the system, and whether this initial depletion layer can be sustained. If double enrichment layers are formed, which is the case for negative voltages, the instability will not appear.

It should be noted that all the results presented in Part I, and those in **Figure 2**, are the steady-state responses of all the various systems considered (unipolar, ideal bipolar, and non-ideal bipolar). In order to calculate the steady-state response (shown in Part I), we first calculated the equilibrium state (i.e. $\langle \bar{i} \rangle = V = 0$), and then at $t = 0$ we applied a nonzero voltage drop across the system $V \neq 0$. In Part II, we shall focus on the time-transient responses, leading up to the more intuitive steady-state response of Part I.

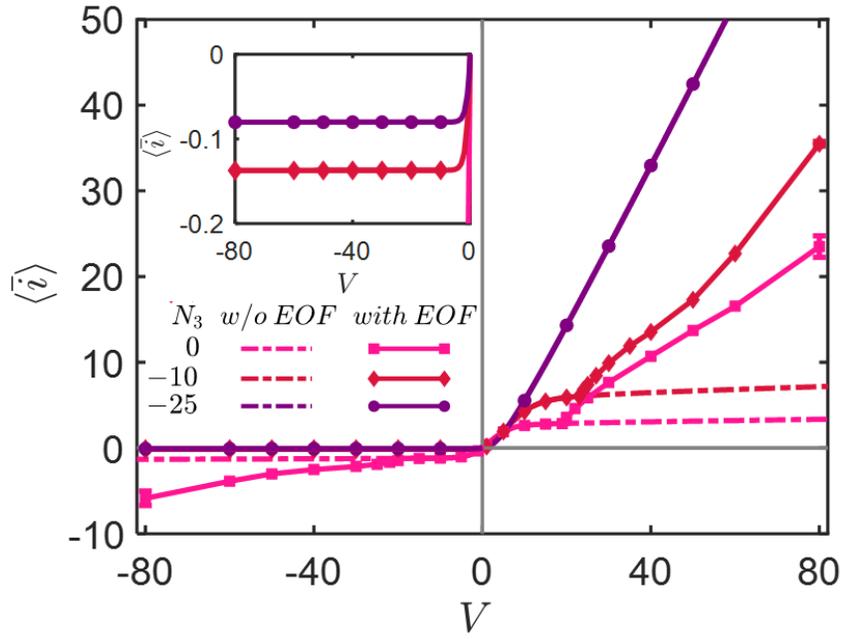

**Figure 2**. Steady-state current density-voltage, $\langle \bar{i} \rangle - V$, results without EOF (dashed lines) and with EOF (solid lines) for three scenarios: unipolar ($N_3 = 0$), non-ideal bipolar ($N_3 = -10$), and ideal bipolar ($N_3 = -25$) systems. The inset is a zoomed view of the negative voltage near $\langle \bar{i} \rangle = 0$, showing that the $\langle \bar{i} \rangle - V$ curves of bipolar systems do not exhibit OLCs there. The error bars denote one standard deviation of the current.

Since the steady-state response and the numerical simulations are already discussed in Part I, in Sec. 2, we only briefly review the theoretical model and numerical simulations (see Part I and its Supplementary Material for a detailed discussion regarding the methods). Thereafter, we divide the paper into three sections that discuss each of the three scenarios depicted in **Figure 2**: unipolar, ideal bipolar, and non-ideal bipolar. We start in Sec. 3 with the time-dependent behavior of the highly investigated unipolar system. While most of the results here have been presented in many past works (references are provided below), a thorough discussion and review are extremely useful as most of the phenomena observed in unipolar systems are observed, in one form or another, in bipolar systems. Thus, the unipolar section "sets the stage" and provides all of the required background needed to understand the bipolar response. In Sec. 4, we move to the other extremity of an "ideal" bipolar response and consider the time-dependent behavior for positive voltages $V > 0$. We shall show that at initial times, two depletion regions are formed, and the EOI appears on both sides of the permselective material. Eventually, due to the symmetry of the system, the EOI decays completely and results in the $\langle \bar{i} \rangle - V$ being unchanged relative to the scenario without EOF. Finally, in Sec. 5, we discuss the transient response of a non-ideal bipolar system and relate this response to the two previous scenarios. For the sake of brevity, the negative voltage, $V < 0$, responses have been moved to the Supplementary Material (SM). In general, the $V < 0$ is uninteresting, as double enrichment layers are formed and EOI cannot appear – such that the response with and without EOF are identical (**Figure 2** inset). Since we provide a preliminary summary for each of the scenarios, the concluding discussion in Sec. 6 is relatively short.

## 2. Problem formulation

We solve the non-dimensional time-dependent equations that govern ion transport through a permselective medium. These are the Poisson-Nernst-Planck and the Stokes equations for a symmetric and binary electrolyte ($z_+ = -z_- = 1$) with ions of equal diffusivities ($\tilde{D}_\pm = \tilde{D}$) in a four-layer system, as shown in **Figure 1**. Note that tilded notations are used for dimensional variables, whereas untilded variables are non-dimensional. Our control parameters are the non-dimensional Debye length, $\varepsilon$, the Péclet number, Pe, the non-dimensional voltage, $V$, and the non-dimensional excess counterion concentration, $N$,

$$\varepsilon = \frac{\tilde{\lambda}_D}{\tilde{L}} = \frac{1}{\tilde{L}}\sqrt{\frac{\tilde{\varepsilon}_0 \varepsilon_r \tilde{\Re} \tilde{T}}{2\tilde{F}^2 \tilde{c}_0}}, \quad \text{Pe} = \frac{\tilde{\varepsilon}_0 \varepsilon_r \tilde{\varphi}_{th}^2}{\tilde{\mu}\tilde{D}}, \quad V = \frac{\tilde{V}}{\tilde{\varphi}_{th}}, \quad N = \frac{\tilde{N}}{\tilde{F}\tilde{c}_0}. \tag{1}$$

Here, $\tilde{\Re}$ is the universal gas constant, $\tilde{T}$ is the absolute temperature, $\tilde{F}$ is the Faraday constant, $\tilde{\varepsilon}_0$ and $\varepsilon_r$ are, respectively, the permittivity of the vacuum and the relative permittivity, $\tilde{c}_0$ is bulk concentration, $\tilde{\mu}$ is the dynamic viscosity of the fluid, and $\tilde{\varphi}_{th} = \tilde{\Re}\tilde{T}/\tilde{F}$ is the thermal potential. The spatial variables have been normalized by a characteristic length $\tilde{L}$ (which can be any of the lengths in the system). The excess counterion is related to the average volumetric space charge density needed to counterbalance the surface charge density. We remind that time, $\tilde{t}$, has been normalized by the diffusion time $\tilde{L}^2/\tilde{D}$, the ionic fluxes have been normalized by $\tilde{j}_0 = \tilde{D}\tilde{c}_0/\tilde{L}$, while the non-dimensional space charge density, $\rho_e$ has been normalized by $\tilde{F}\tilde{c}_0$. The non-dimensional velocity vector $\mathbf{u} = u\,\hat{x} + v\,\hat{y}$ and pressure $p$ have been normalized, respectively, by a typical velocity $\tilde{u}_0 = (\tilde{\varepsilon}_0 \varepsilon_r \tilde{\varphi}_{th}^2)/(\tilde{\mu}\tilde{L})$ and pressure $\tilde{p}_0 = \tilde{\mu}\tilde{u}_0/\tilde{L}$.

Also, it is important to note that we have chosen to work in a non-dimensional formulation to reduce the number of parameters to a minimal number. In particular, in all of our simulations, we keep $\varepsilon$ and Pe constant while we vary $V$ and $N$ (three tables of all simulation parameters are given in the supplementary material of Part I). Note that a constant non-dimensional number doesn't imply that all parameters are not varied but rather that a multiplication of all the relevant parameters is not varied. As such, performing the simulation in a non-dimensional manner is more robust than performing them in a dimensional manner.

Our two-dimensional (2D) system consists of either three or four regions (always two diffusion layers, with one or two permselective regions) representing unipolar or bipolar systems, respectively. A detailed discussion on the boundary conditions is given in Part I, and it will not be repeated here. We remind the reader that in the permselective regions, we don't account for hydrodynamic effects. Finally, we remind that we will use the $\Delta_k$ notations to denote cumulative lengths within the system such that $\Delta_1 = L_1$, $\Delta_2 = \Delta_1 + L_2$, $\Delta_3 = \Delta_2 + L_3$, $\Delta_4 = \Delta_3 + L_4$.

Throughout, we will use the following (1D and 2D) spatial (denoted with single and double overbars, respectively) and temporal (denoted with chevron brackets) averaging operators of any quantity $f$ (e.g., $c_\pm$, $\varphi$, $\rho_e$, and their fluxes, etc.) defined by

$$\bar{f}(y,t) = \frac{1}{W}\int_0^W f(x,y,t)dx, \quad \bar{\bar{f}}_{k=1,4}(t) = \frac{1}{L_{k=1,4}}\int_{L_{k=1,4}} \bar{f}(y,t)dy, \tag{2}$$

$$\langle \bar{f} \rangle = \frac{1}{T}\int_{t_0}^{t_0+T} \bar{f}\,dt, \quad \langle \bar{\bar{f}} \rangle = \frac{1}{T}\int_{t_0}^{t_0+T} \bar{\bar{f}}\,dt, \tag{3}$$

where, $t_0$ is the time at which the system reaches a state where the state is perturbed about a "steady-state" (i.e., a statistic steady-state), and $T$ is the time duration of this "steady-state" (typically the end of the simulations). Of particular importance are the averages for the non-dimensional normal electrical current density (normalized by $\tilde{F}\tilde{D}\tilde{c}_0/\tilde{L}$), $i_y(t)$, at $y=0$ given by $\bar{i}(t) = W^{-1}\int_0^W i_y(x, y=0, t)dx$ and the non-dimensional kinetic energy density (normalized by $\tilde{\rho}\tilde{u}_0^2$) $E_k = \frac{1}{2}|\mathbf{u}|^2 = \frac{1}{2}(u^2+v^2)$. In the following, we will present an analysis of the time-dependent behavior of the 1D averages of the cation concentration, $\bar{c}_+$, the space charge density, $\bar{\rho}_e$, and kinetic energy $\overline{E_k}$ in Regions 1 and 4.

Similar to Part I, which focuses on the steady-state $\langle \bar{i} \rangle - V$, here, we will consider the time transient response of three scenarios: unipolar, non-ideal bipolar, and ideal bipolar. All these systems are characterized by the ratio of the geometry and excess counterion charge density of both permselective regions [see Green, Edri, and Yossifon (2015) for the 2D version of this equation]

$$\eta = \frac{L_3}{L_2} \times \left|\frac{N_3}{N_2}\right| = \begin{cases} 0 & , \text{ unipolar} \\ 0 < \eta < 1, & \text{non-ideal bipolar} \\ 1 & , \text{ ideal bipolar} \end{cases} \quad (4)$$

It is trivial to see that when either $N_3 = 0$ or $L_3 = 0$, the second charged region does not exist, and the bipolar system reduces to the unipolar system. If, however, there are two charged regions, the ratio $\eta$ will determine the overall response of the system. If $\eta = 1$ (and $L_2 = L_3$), the total excess counterion charges in both regions are equal (but of opposite sign), and the response is what we now term 'ideal' bipolar. If $1 > \eta > 0$, we term the response non-ideal bipolar. The non-ideal bipolar system will exhibit time-dependent and steady-state characteristics of both unipolar and ideal bipolar.

## 3. Time-dependent response of a unipolar system

This section considers the scenario of a unipolar system with only one permselective region. The unipolar system has been investigated both in a one-layer system with a perfect permselective surface (Rubinstein & Zaltzman, 2000; Zaltzman & Rubinstein, 2007; Rubinstein et al., 2008; Chang, Yossifon, & Demekhin, 2012; Pham, Li, Lim, White, & Han, 2012; Demekhin, Nikitin, & Shelistov, 2013; Druzgalski, Andersen, & Mani, 2013; Deng et al., 2013; de Valença, Wagterveld, Lammertink, & Tsai, 2015; Mani & Wang, 2020; Sensale, Ramshani, Senapati, & Chang, 2021; Zhang, Zhang, Luo, Yi, & Wu, 2022; Pandey & Bhattacharyya, 2022; Chen, Zhang, Zhang, Luo, & Yi, 2023) and a three-layer system with symmetric diffusion layers, $L_1 = L_4$ (Kim, Wang, Lee, Jang, & Han, 2007; Yossifon & Chang, 2008; Rubinstein & Zaltzman, 2015; Abu-Rjal, Rubinstein, & Zaltzman, 2016; Abu-Rjal, Prigozhin, Rubinstein, & Zaltzman, 2017; Demekhin, Ganchenko, & Kalaydin, 2018). Without loss of generality, we set $N_3 = 0$ such that the four-layer system (**Figure 1**) is reduced to a three-layer unipolar system in which a highly cation-permselective medium is flanked by two diffusion layers [**Figure 3**(a)]. Here, we will investigate a three-layer system with *asymmetric* diffusion layers[1] for the particular case that $L_4 = 2L_1$.

**Figure 3** and **Figure 4** show the time-transient response of a three-layer highly permselective system with asymmetric diffusion layers at a positive OLC regime. **Figure 3**(b) plots $\bar{i}(t)$ for two scenarios: without EOF (dashed lines) and with EOF (solid lines). Without EOF, it can be observed that the current density decreases monotonically (Abu-Rjal et al., 2019) [this is also true for a "symmetric" three-layered system without EOF [Supplementary Material (SM), **Figure S1**]. In contrast, for the scenario with EOF, the change is highly non-monotonic. This can be attributed to the appearance of the EOI (the time evolution is shown in **Figure 4** and **Movies 1-3** in SM), which is also responsible for increasing $\bar{i}(t)$ (relative to the scenario without EOF).

At $t = 0$, the system starts at equilibrium; this is tantamount to setting the concentrations to unity, $c_\pm(t=0) = 1$, and $\rho_e(t=0) = 0$ everywhere outside of the equilibrium EDL. Upon application of a positive voltage, $V > 0$, it can be observed from **Figure 3**(c) that outside the quasi-equilibrium EDL, the concentration in Region 1 is always smaller than unity, while in Region 4 [**Figure 3**(d)], the concentration is larger than unity (see **Movie 1** in the SM). Naturally, these regions are called the depleted and enriched regions, respectively, and are an essential component of concentration polarization [see Levich (1962), Rubinstein & Zaltzman (2000), Zaltzman & Rubinstein (2007), Chang, Yossifon & Demekhin (2012) and Mani & Wang (2020) for thorough discussions on concentration polarization].

In the enriched region (Region 4), the large concentrations [**Figure 3**(d)] ensure that $\bar{\rho}_e$ [**Figure 3**(f)] maintains its 1D quasi-equilibrium structure such that EOI doesn't appear and the average kinetic energy is zero at all times [**Figure 3**(h)]. As will be discussed further below, it is essential to realize and remember that enriched regions cannot support nonequilibrium ESCs and that the ESC can only form on the depleted side when the concentration at the interface ($y = \Delta_1$) approaches zero.

Naturally, the dynamics in the depleted side are by far more complicated. Around $t = 0.011$, the concentration at the interface ($y = \Delta_1$) approaches zero [**Figure 3**(c)]. As a result, the space charge density moves from a quasi-equilibrium to a nonequilibrium structure, which is commonly known as the extended-space charge (ESC) [**Figure 3**(e)] (Rubinstein & Shtilman, 1979; Zaltzman & Rubinstein, 2007; Yariv, 2009). This nonequilibrium structure is unstable to lateral perturbations (Rubinstein & Zaltzman, 2000; Zaltzman & Rubinstein, 2007; Kim et al., 2007; Rubinstein et al., 2008; Yossifon & Chang, 2008; Pham et al., 2012; Mani & Wang, 2020), and eventually, the EOI forms [**Figure 4** and **Movie 2**]. As a result, the average kinetic energy, $\bar{E}_k$, also increases such that the effects of the velocity span the entire depleted region [**Figure 3**(g)] (Demekhin, Nikitin, & Shelistov, 2013; Druzgalski, Andersen, & Mani, 2013).

---

[1] It is essential to note that our definition of symmetry and asymmetry in the unipolar system is regarding the diffusion layers. In contrast, for bipolar systems, we deal with a completely different kind of "symmetry/asymmetry" discussed in Part I.

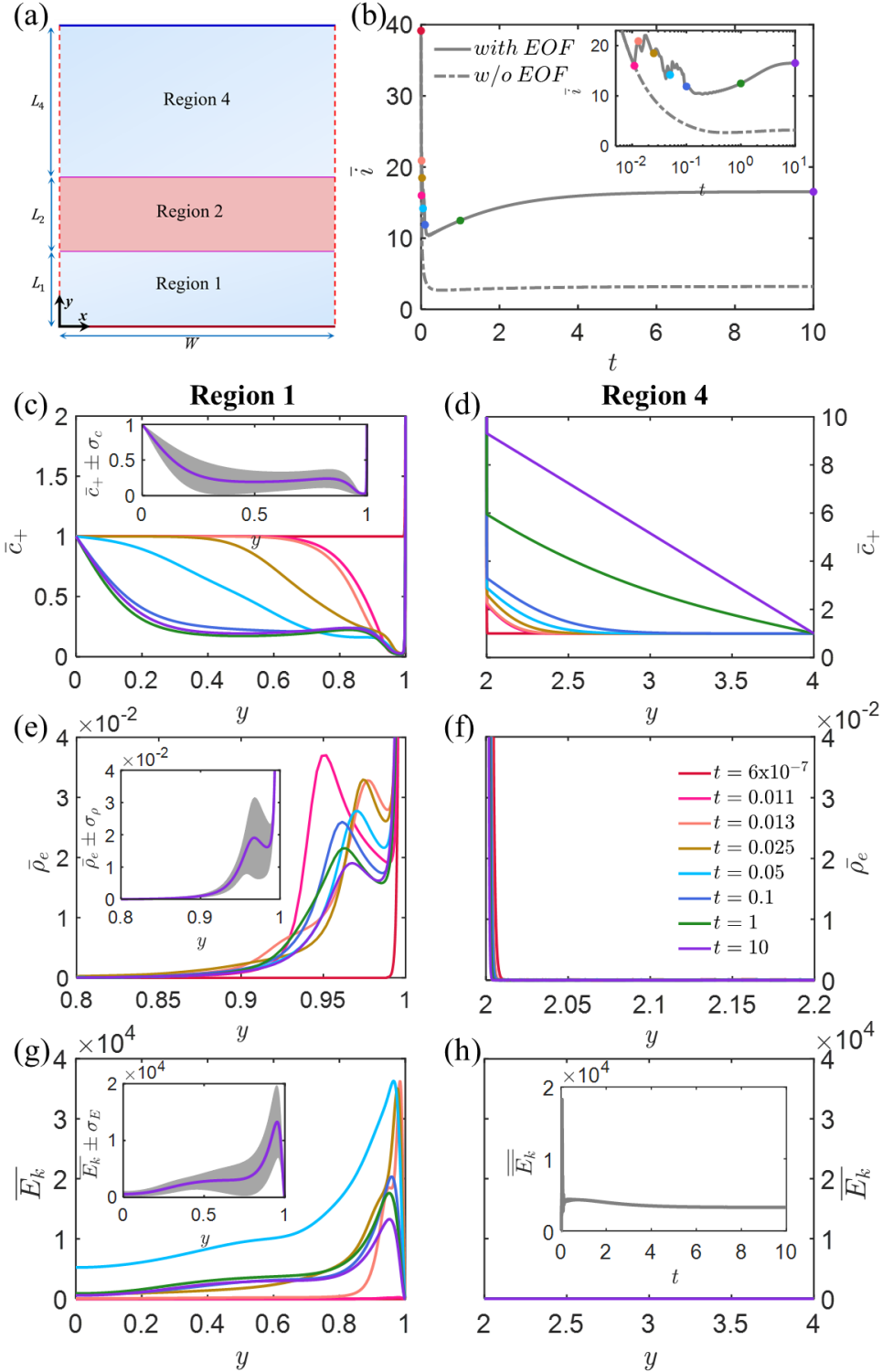

*Figure 3*. (a) Schematic describing a 2D three-layered system comprised of two asymmetric diffusion layers (Regions 1 and 4) connected by a <u>single highly</u> permselective medium (Region 2). (b) The electric current density versus time, $\bar{i}(t)$, response of the system with and without EOF. The inset of (b) shows a zoom-up on a semilog$_{10}$ plot. (c-h) The spatial-averaged time-dependent profiles of (c and d) the concentration $\bar{c}_+$, (e and f) the space charge density $\bar{\rho}_e$, and (g and h)

the kinetic energy $\overline{\overline{E_k}}$ in Regions 1 and 4 [based on the first equation of Eq.(2)]. The colors for each line correspond to the markers shown in (b) and the legend is given in (f). The insets in (c), (e), and (g) show profiles for the last time-point for $c_+$, $\rho_e$, and $E_k$, respectively, while the shaded grey regions denote the standard deviation of each variable respectively ($\sigma_c$, $\sigma_\rho$ and $\sigma_E$). The inset in (h) is the time evolution of the surface averaged kinetic energy, $\overline{\overline{E_k}}(t)$, in <u>Region 1</u> [based on the second equation of Eq.(2)]. This figure uses $V = 60$.

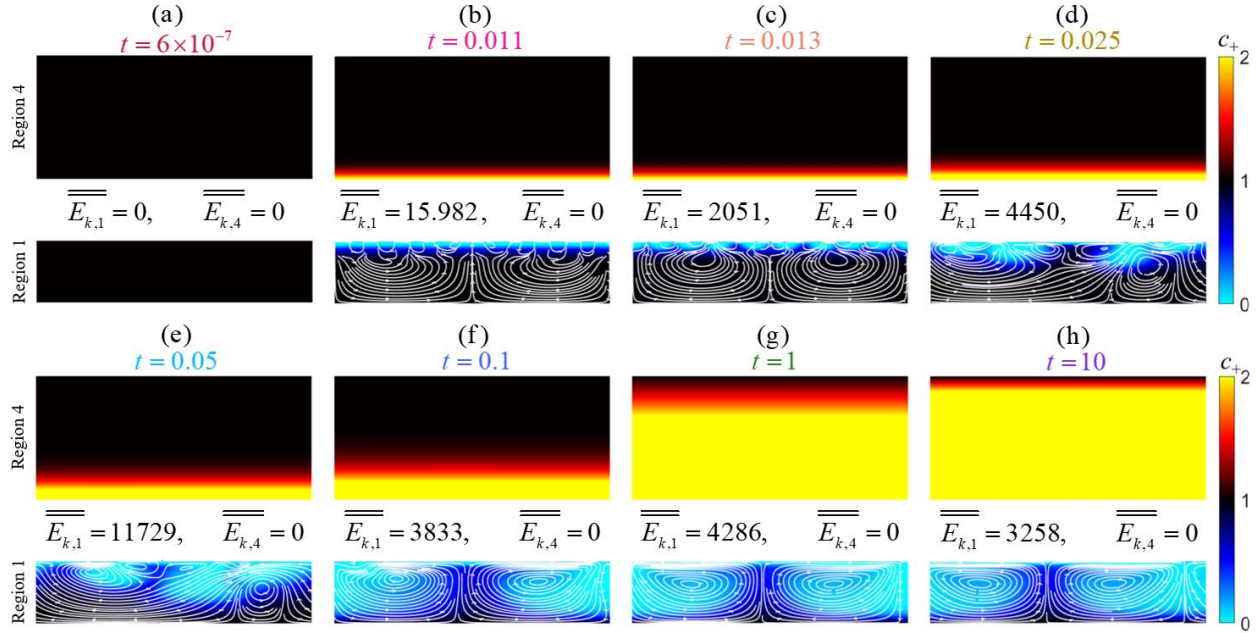

**Figure 4**: *Time evolution of the positive concentrations (2D color plot), $c_+$, and velocity streamlines (white lines) in Regions 1 and 4 for the simulation of **Figure 3**. The colored titles of each sub-figures follow the color notation of **Figure 3**. The surface average of the kinetic energy, $\overline{\overline{E_k}}$ [based on the second equation of Eq.(2)], is given for each snapshot (subscripts of numbers denote regions).*

**Preliminary summary.** In this section, we have revisited the highly investigated time-dependent response for a unipolar system. We have shown that upon application of a positive voltage (negative voltages are discussed in the SM), a depleted layer and an enriched layer form in each of the diffusion layers. This is in contrast to the bipolar systems, which at early times will have two depleted layers and at later times have no depleted and enriched layers (ideal bipolar) or a depleted and an enriched layer (non-ideal bipolar). Above a critical voltage, an EOI forms only on the depleted sides. The EOI, which introduces and injects kinetic energy into the system, leads to over-limiting currents (OLCs). Importantly, so long as a supercritical voltage is applied, the EOI does not decay. This stands in contrast to the dynamics that will be observed for bipolar systems.

Finally, for the sake of completion and brevity, we note that the unipolar system differs from the bipolar systems in that the EOI can form in either diffusion layer if the positive voltage is changed to negative voltage. Also, we note that breaking the geometric symmetry of the diffusion layers leads to quantitatively different responses that are without any qualitative change. All these are shown in the SM.

## 4. Time-dependent response of an ideal bipolar system

This section focuses on an ideal bipolar system where the system satisfies $\eta = 1$ (i.e., $N_3 = -25$). We will show that the "ideal" bipolar system has completely different characteristics compared to that of the unipolar system (Sec.3). The layout of **Figure 5** and **Figure 6** are similar to those of **Figure 3** and **Figure 4**, respectively, save that we now consider an ideal bipolar four-layer setup [**Figure 5**(a)].

**Figure 5**(b) plots $\bar{i}(t)$ for two scenarios: without EOF (dashed lines) and with EOF (solid lines). The scenario without EOF shows a sharp decrease and then a monotonic rise to the predicted steady-state current [Green, Edri & Yossifon (2015), and Abu-Rjal & Green (2021) provide the expression for the convection-less $\langle \bar{i} \rangle - V$ - which extends the $\langle \bar{i} \rangle - V$ of Vlassiouk et al. (2008), from 1D to 2 and accounts for the resistances of the diffusion layers]. The scenario with EOF shows a similar drop and rise with several remarkable features. First, unsurprisingly, when accounting for EOF, the EOI appears (**Figure 6** and **Movies 4-6**). However, surprisingly, the EOI appears in both diffusion layers. Second, with the appearance of EOI, we see a relative increase of $\bar{i}(t)$ (relative to the scenario without EOF). This, too, is unsurprising since EOF is more efficient than diffusion in stirring the electrolyte. Third, since EOI is an unstable process, it is natural that the changes are non-monotonic and "noisy" [inset of **Figure 5**(b)]. Fourth, surprisingly, the steady state current with EOF equals the current of the no-EOF scenario. To understand this last point, it is essential to understand the behavior of $\bar{c}_\pm$, and $\bar{\rho}_e$ in Regions 1 and 4.

At $t = 0$, the system starts at equilibrium. For the concentrations, this is tantamount to $c_\pm(t = 0) = 1$ and $\rho_e(t = 0) = 0$ everywhere outside of the equilibrium EDL. **Figure 5**(c)-(d) show the behavior of $\bar{c}_+$ in Regions 1 and 4, respectively, while **Figure 5**(e)-(f) show the behavior of $\bar{\rho}_e$ in Regions 1 and 4, respectively. In contrast to the unipolar case considered in **Figure 3**, where one of the diffusion layers is depleted and the other is enriched, here we observe much more complicated dynamics – including double diffusion layers and much more.

**Figure 5**(c)-(d) show that both diffusion layers become depleted. **Figure 5**(e)-(f)] show that in both layers, an ESC layer is formed (in the unipolar system, only one ESC was formed – in the single depletion layer). Equally remarkable, the dynamics of the concentration fields are symmetric (for the unipolar case, there were both depleted and enriched layers), while the space charge density fields are antisymmetric ($\bar{\rho}_{e,1} = -\bar{\rho}_{e,4}$). At initial times, the effects of depletion increase such that the interfacial concentrations decrease, but the space charge still has an equilibrium structure ($t < 0.004$). At intermediate times ($t \sim 0.004$), a nonequilibrium space charge region is formed [pink lines **Figure 5**(c)-(f)]. Since the nonequilibrium space charge region is unstable, EOI appears in <u>both</u> regions! For times larger than $t > 3$, the effect of EOI appears to decay. The reason for this is explained below. Finally, at steady-state after the EOI has completely decayed [best observed by the decay of the kinetic energy in **Figure 5**(g)-(h)], $\bar{c}_\pm$ and $\bar{\rho}_e$ return to their quiescent states.

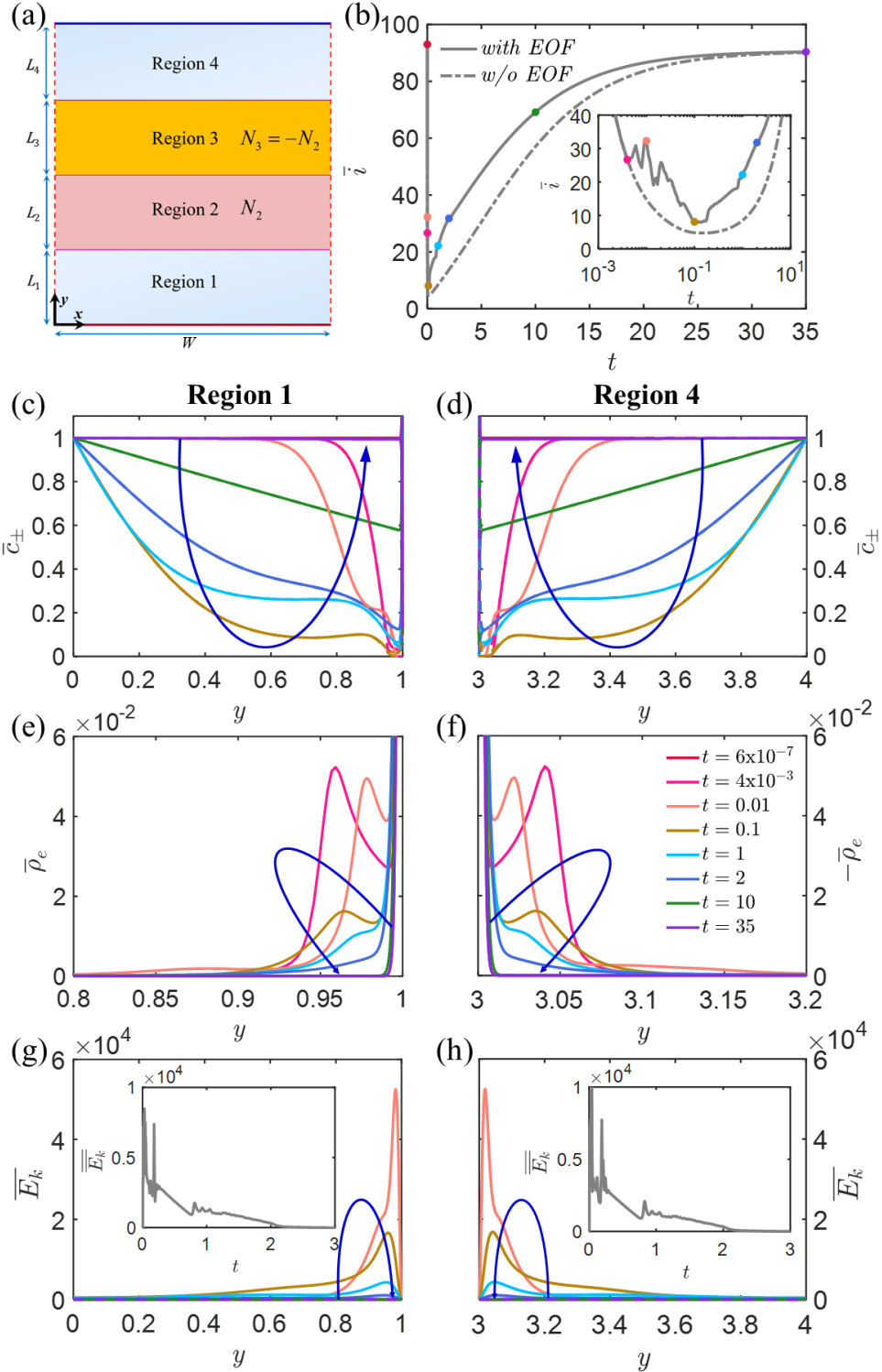

***Figure 5***. *(a) Schematic describing a 2D four-layered ideal bipolar system comprised of two diffusion layers (Regions 1 and 4) connected by two permselective mediums (Regions 2 and 3). (b) The electric current density versus time, $\bar{i}(t)$, response of the system with and without EOF. The inset of (b) shows a zoom-up of early times on a semilog$_{10}$ plot. (c-h) The spatial-averaged time-dependent profiles of (c and d) the concentration $\bar{c}_+$, (e and f) the space charge density $\bar{\rho}_e$, and*

*(g and h) the kinetic energy $\overline{\overline{E_k}}$ in Regions 1 and 4. Note that in (e), we present $\overline{\rho}_e$, while in (f), we present $-\overline{\rho}_e$. The colors for each line correspond to the markers shown in (b) and the legend is given in (f). The insets in (g) and (h) are the time evolution of the surface average of the kinetic energy, $\overline{\overline{E_k}}(t)$, in Regions 1 and 4, respectively. The blue curved arrows indicate the direction of increasing time, showing the non-monotonic response. Here, we have used $V = 100$.*

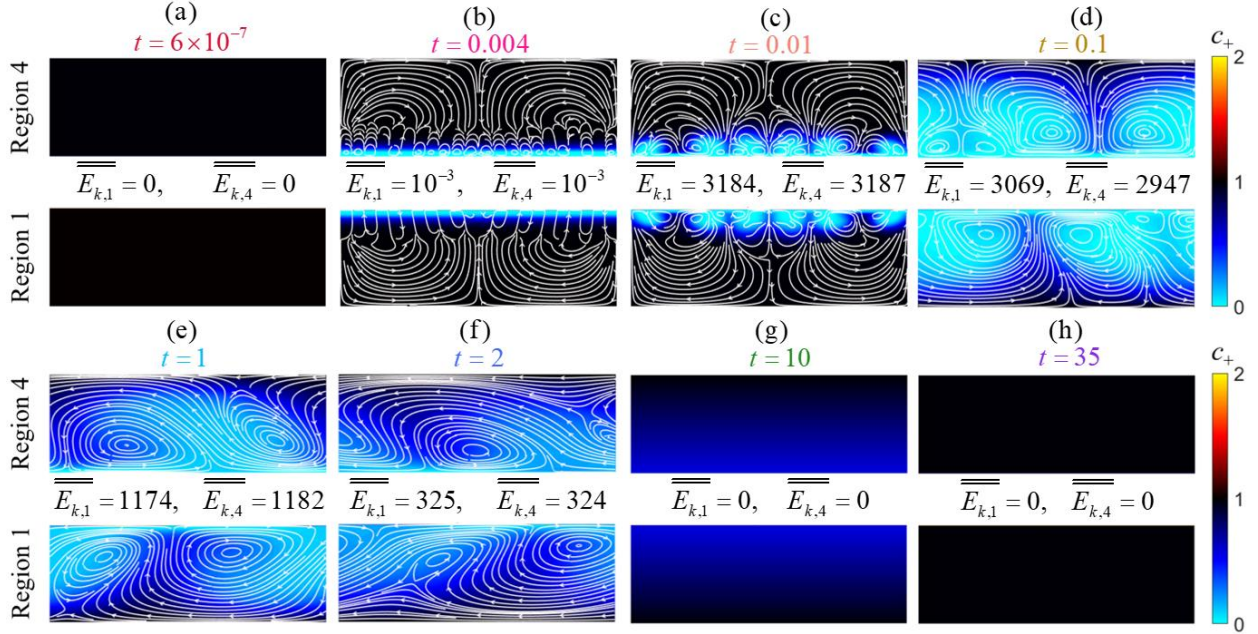

*Figure 6. Time evolution of the positive concentrations (2D color plot), $c_+$, and velocity streamlines (white lines) in Regions 1 and 4 for the simulation of Figure 5. The colored titles of each sub-figures follow the color notation of Figure 5. The surface average of the kinetic energy, $\overline{\overline{E_k}}$, is given for each snapshot (subscripts of numbers denote regions).*

**Movie 4** in the SM plots the time-dependent behavior of the distribution of $\overline{c}_+$ and $\overline{\rho}_e$ for the two scenarios: without and with EOF. Except for the appearance of EOI and its eventual decay, the qualitative nature of the EOF results is almost qualitatively identical to the response of an ideal convection-less bipolar system (Abu-Rjal & Green, 2021; Tepermeister & Silberstein, 2023). We found this result somewhat surprising, leading to the question we alluded to in Part I (Abu-Rjal & Green, 2024) – why are the steady-state responses without and with EOF identical?

The answer is surprisingly simple and is best understood by considering the ideal convection-less bipolar system for the particular case $\eta = 1$. In our past work [Green, Edri, & Yossifon, 2015], we showed that one can derive an analytical $\langle \overline{i} \rangle - V$ for the case of $\eta = 1$ if one assumes that the salt current density, $j = j_+ + j_-$, is zero ($j = 0$). In this scenario, positive ions are transported via $j_+$ from Region 1 to Region 4 through the bipolar region. During the time they are transported to the interface of Regions 3 and 4, the (negative) ESC has formed. The arrival of the positive ions stabilizes the ESC by reducing its magnitude. A similar process occurs for negative ions via $j_-$.

By virtue of symmetry but opposite directions, $j_+ = -j_-$, the response is symmetric in both Regions 1 and 4.

The time-dependent dynamics are a bit more complicated; however, the simplicity of the explanation remains. At $t=0$, prior to the application of the voltage drop, $V$, the fluxes are zero such that $i = j = 0$. As stated above, at steady state, there is still a constraint that $j=0$. However, in between $t=0$ and $t \to \infty$, $j(t)$ varies locally (and globally) such that it increases before going back to zero. Since $j$ and $i$ are linked, this also leads to changes in $i$. The local behavior can be observed in SM **Movie 6**, while **Figure S4** shows the correlation between $\bar{\bar{j}}(t)$ in Region 1 and $\bar{i}(t)$ for both scenarios (without and with convection). It can be observed that the non-monotonic changes in $\bar{i}(t)$ are related to the non-monotonic changes in $\bar{\bar{j}}(t)$ due to the instability. A thorough discussion of the time-dependent dynamics of all the fluxes, without the EOF, is given in (Abu-Rjal & Green, 2021).

For the current scenario of an ideal bipolar system with EOF, one must return to the problem definition in Sec. 2 [and Part I Sec. 3.2], where we assumed that the velocity within the permselective material is zero. As a result, the response of the bipolar membrane (Regions 2 and 3) must remain unchanged relative to the response of the bipolar membrane in the convectionless scenario. Here, too, positive ions are transported via $j_+$ from Region 1 to Region 4 in such a manner as the ESC in Region 4 is stabilized. Naturally, the exact same thing occurs for negative ions via $j_-$ which stabilize the positive ESC in Region 1. In this manner, both ESCs are stabilized and diminished such that the electric body force that drives the EOF instability is diminished, and the instability decays – this can be observed in the decay of the kinetic energy [**Figure 5**(g)-(h)].

Before summarizing this section, three last comments are needed. First, we start with a technical comment. In general, for the ideal bipolar scenario, the response in both diffusion layers should have a mirror reflection (and possibly a sign difference) – this is what occurs for a bipolar system without EOI. However, in several time steps in **Figure 6**, the $\bar{\bar{E_k}}$ in Regions 1 and 4 are not identical. Since the spatial averages of the concentrations [**Figure 5**(c)-(d)] and space charge densities [**Figure 5**(e)-(f)] are the same, as are the averages for the velocities, we believe this few percent error can be attributed to meshing and the fact that we are considering the square of the velocities (where any mismatch is drastically enhanced). Second, we point out that the appearance and decay of the EOI do not occur for a specific voltage. Rather, it occurs for all the voltages we considered. In the SM, we demonstrate this statement for several voltages. Third, for an ideal bipolar system subject to negative voltage drops, $V < 0$, the EOI doesn't appear at all. This is because for $V < 0$, double enrichment layers, that do not support an ESC, forms. This, too, is demonstrated in the SM [and discussed thoroughly in our past work (Abu-Rjal & Green, 2021)].

**Preliminary summary.** In summary, the response of the ideal bipolar system differs substantially from the unipolar system in that either double depletion (for $V > 0$) or double enrichment (for $V < 0$) layers are formed. When double depletion layers are formed, the EOI appears in both layers. However, in contrast to the unipolar, even when an over-critical voltage is applied, the EOI decays such that the steady-state over-limiting current cannot be observed. Importantly, the time-dependent response of the system with EOI includes a clear signature that can be observed relative to the rather quiescent scenario without EOI [**Figure 5**(b)].

## 5. Time-dependent response of a non-ideal bipolar system

This section considers the non-ideal bipolar case, $1 > \eta > 0$, [**Figure 7**(a)]. We will show that the response of this system displays some characteristics observed in the ideal bipolar case (Sec. 4) and characteristics observed in the unipolar system (Sec. 3). To highlight the main shared features and differences, **Figure 7** and **Figure 8** have the same layout of **Figure 3** and **Figure 4** in Sec. 3 and **Figure 5** and **Figure 6** in Sec. 4.

**Figure 7**(b) plots $\bar{i}(t)$ for the two scenarios of without EOF (dashed lines) and with EOF (solid lines). The scenario without EOF shows a sharp decrease and then a monotonic rise to the steady-state current. Here, too (similar to the $\eta = 1$ scenario), the non-monotonic change of the current should also be attributed to the non-monotonic change of $\bar{\bar{j}}(t)$ (**Figure S7**), which shows that $\bar{\bar{j}}(t=0) = 0$ but changes until it reaches its steady-state value. Similarly, the scenario with EOF exhibits a non-monotonic change similar to the unipolar and ideal bipolar scenarios. This can be related to the appearance of EOI (**Figure 8** and **Movies 7-9**). Importantly, in contrast to the ideal bipolar scenario [**Figure 5**(b)] where the steady-state currents without EOF and with EOF were identical, here we observe that the steady-state current without EOF is substantially smaller than that with EOF – as was observed in the unipolar scenario [**Figure 3**(b)].

To better understand the change in the behavior of $\bar{i}(t)$, we consider the behavior of $\bar{c}_+$, and $\bar{\rho}_e$ in Regions 1 and 4. Once more, at $t = 0$, the system starts at equilibrium [$c_\pm(t=0) = 1$ and $\rho_e(t=0) = 0$ everywhere outside of the equilibrium EDL]. The $t > 0$ results show a combination of both unipolar and ideal bipolar behaviors:

*Concentrations.* **Figure 7**(c) shows that the depletion layer in Region 1 forms almost monotonically, and the concentration doesn't return to its equilibrium state [$c_+(t=0) = 1$]. This behavior is reminiscent of the unipolar response. The behavior of the concentration in Region 4 [**Figure 7**(d)], is by far more complicated. Similar to the ideal bipolar scenario, the interfacial concentration decreases to form a depletion layer. Once the depletion layer has achieved a minimal value, a reversal is observed. Then, in contrast to the ideal bipolar scenario and more reminiscent of the unipolar response, the concentration does not return to its equilibrium state [$c_+(t=0) = 1$] and an enrichment layer is formed.

*Space charge density.* The behavior of the space charge density [**Figure 7**(e)-(f)] follows the behavior of the concentrations in that the dynamics are by far more complicated. In particular, we observe that $\bar{\rho}_{e,1} \neq -\bar{\rho}_{e,4}$. The ESC in Region 1 [**Figure 7**(e)], whose dynamics are slightly non-monotonic, does not decay at later times (this is reminiscent of the unipolar response). The space charge density in Region 4 [**Figure 7**(f)] mirrors the dynamics of the concentrations: when the layer is depleted, an ESC forms; when the layer becomes enriched, the ESC disappears.

*Kinetic energy.* Naturally, the behavior of the velocity must follow that of the space charge density (**Figure 8**). As in the ideal bipolar scenario, EOI appears in both Regions 1 and 4. In contrast to the ideal bipolar scenario and similar to the unipolar scenario, the EOI in Region 1 doesn't decay [**Figure 7**(g)]. Similar to the ideal bipolar scenario and the unipolar scenario, the EOI in Region 4 decays [**Figure 7**(h)]. Importantly, the permanence of the EOI in Region 1 leads to persistent OLC in the non-ideal bipolar case [**Figure 7**(b)].

It should be noted that since $1 > \eta > 0$, there is an inherent asymmetry between $j_+$ and $j_-$ fluxes. In particular, we note that the $j_+$ fluxes, leaving Region 1 and arriving in Region 4, deplete the ESC in Region 1 quicker than the $j_-$ fluxes, leaving Region 4 and arriving in Region 1, can stabilize the ESC. In contrast, it is easy to observe that the $j_+$ fluxes stabilize the ESC in Region 4 quicker than the depletion of $j_-$ fluxes. As a result, the EOF can be observed at steady-state in Region 1 while it is not observed in Region 4. The difference in the behavior of $j_+$ and $j_-$ in Regions 1 and 4 also leads to differences in the behavior $\bar{\bar{j}}$ and $\bar{i}$ in these regions. Similar to **Figure S4** shown in Sec. 2, **Figure S7** presents the time-dependent of these quantities, showing that differences in the behavior of $\bar{\bar{j}}_1(t)$ and $\bar{\bar{j}}_4(t)$, which are also responsible for the non-monotonic change in $\bar{i}(t)$. See **Movie 9** for more details.

To complete our analysis of non-ideal bipolar systems, we need two more things: more positive voltages and negative voltages. For the sake of completion, in the SM, we provide several $\bar{i}(t)$ curves for several additional positive voltages where we demonstrate that the overall characteristics remain the same with the sole difference that when the velocity field becomes chaotic, so does $\bar{i}(t)$. Here, similar to the ideal bipolar response, for negative applied voltages, both diffusion layers exhibit enrichment at early times. At later times, the small limiting currents, determined by the permselective regions (and by the diffusion layers, as in the case of the unipolar system), ensure that the concentration gradients are extremely small, and the space charge density is always in a quasi-equilibrium-like structure. Hence, instability is not observed.

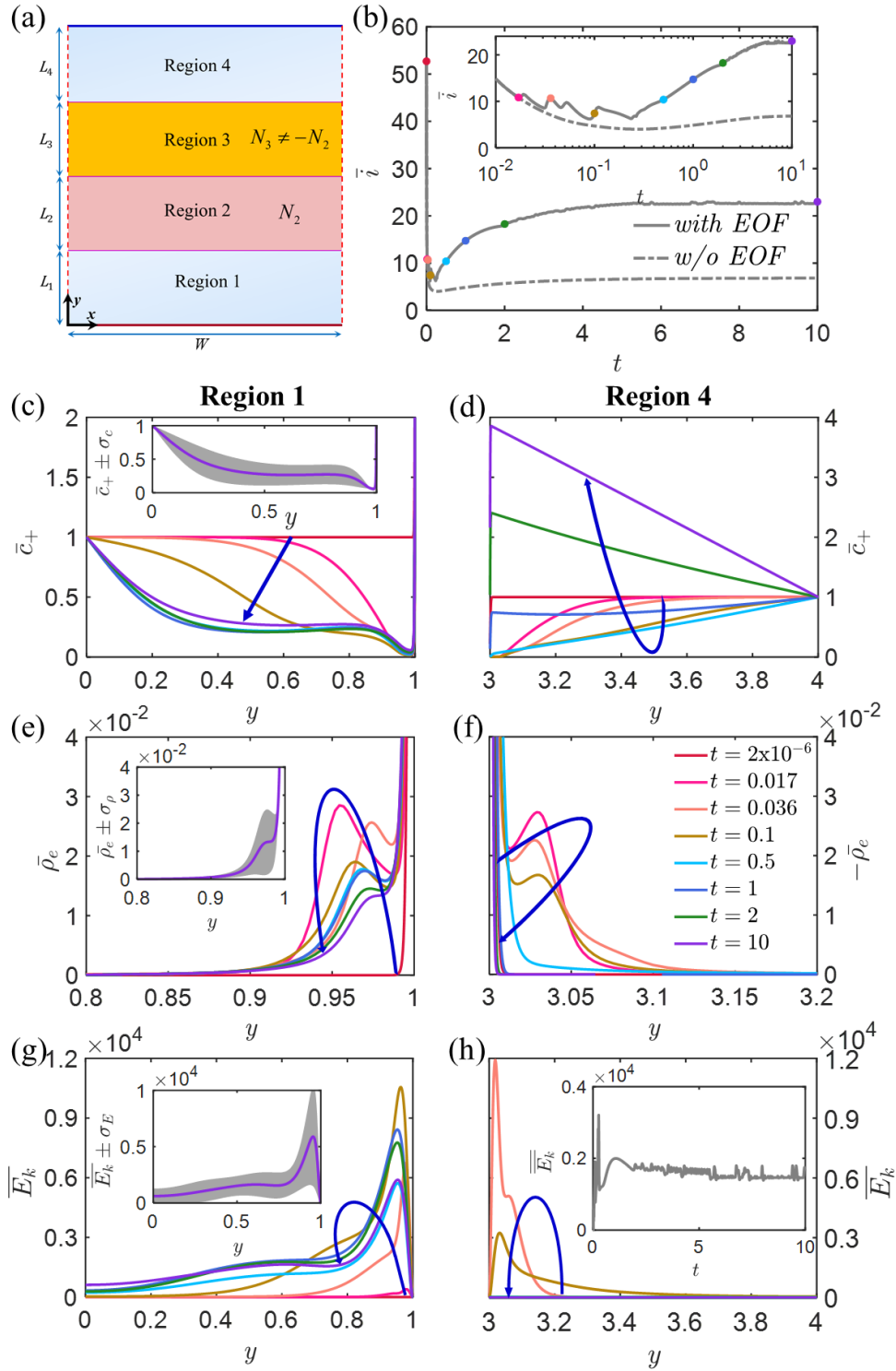

***Figure 7.*** *(a) Schematic describing a 2D four-layered non-ideal bipolar system comprised of two diffusion layers (Regions 1 and 4) connected by two permselective mediums (Regions 2 and 3). (b) The electric current density versus the time, $\bar{\bar{i}}(t)$, response of the system with and without EOF. The inset of (b) shows a zoom-up on a semilog$_{10}$ plot. (c-h) The spatial-averaged time-dependent profiles of (c and d) the concentration $\bar{c}_+$, (e and f) the space charge density $\bar{\rho}_e$, and (g and h) the*

kinetic energy $\overline{\overline{E_k}}$ in Regions 1 and 4. The colors for each line correspond to the markers shown in (b), and the legend is given in (f). The insets in (c), (e), and (g) show the profiles for the last time-point for $c_+$, $\rho_e$, and $E_k$, respectively, while the shaded grey regions denote the standard deviation of each variable respectively (i.e., $\sigma_c$, $\sigma_\rho$ and $\sigma_E$). The inset in (h) is the time evolution of the surface averaged kinetic energy, $\overline{\overline{E_k}}(t)$, in <u>Regions 1</u>. The blue curved arrows indicate the direction of increasing time. Here, we have used $N_3 = -10$ and $V = 60$.

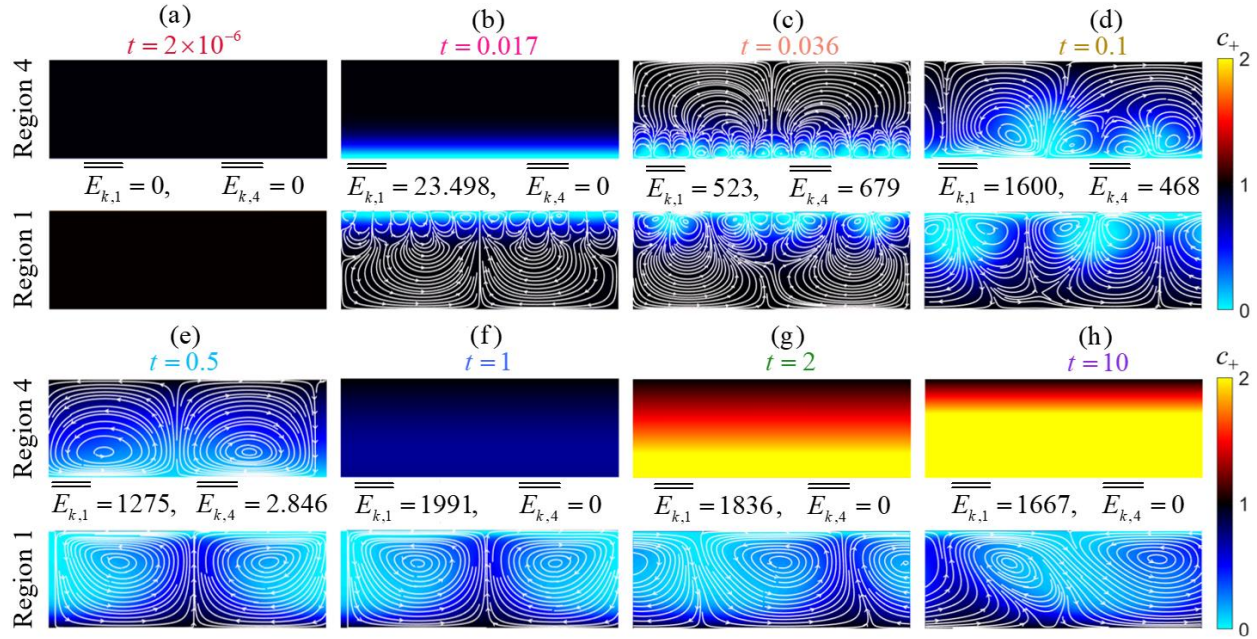

***Figure 8***. *Time evolution of the positive concentrations (2D color plot), $c_+$, and velocity streamlines (white lines) in Regions 1 and 4 for the simulation of **Figure 7**. The colored titles of each sub-figure follow the color notation of **Figure 7**. The surface average of the kinetic energy, $\overline{\overline{E_k}}$, is given for each snapshot (subscripts of numbers denote regions).*

**Preliminary summary.** Here, we have shown that the response of the non-ideal bipolar scenario includes many characteristics that can be attributed to either the unipolar response or the ideal bipolar response (e.g., two time-transient depletion layers, which can be found in the ideal bipolar system, but one depletion and one enrichment layer found in the steady-state response of the unipolar system). The reason that the depletion layers continue to be sustained in one of the regions – is because there is a symmetry breaking between $j_+$ and $j_-$, due to $\eta \neq 1$, leading to both $j \neq 0$ and at least one persistent depletion layer. Importantly, so long as there is a single depletion layer (that is completely depleted), there will also be a nonequilibrium space charge layer that can sustain the EOI (which leads to over-limiting currents).

Admittedly, while our results have provided new insights into the behavior of the EOI in bipolar systems, many questions remain unanswered. Notably, the conditions that lead to the instability and the various scaling laws observed in Part I, are not understood. To that end, a thorough linear stability analysis is needed to determine the effects of the control parameters given in Eq. (1). However, this is left for future work.

## 6. Discussion and summary

This work focuses on elucidating the behavior of the time-transient response of ideal and non-ideal bipolar systems subjected to EOI (Sec. 4 and 5). To that end, we simulated several ideal and non-ideal bipolar systems for a wide range of $V$ and $\eta$. To highlight the surprising and remarkable results uncovered in our simulations of the bipolar system, we first leveraged our understanding of the time-transient response of the more investigated unipolar system (Sec. 3). There it has been known, for quite some time, that the EOI is responsible for OLCs.

In ideal bipolar systems, regardless of whether EOF is included, double depletion layers are formed. Whether EOI forms or not, at steady-state, the system returns to its equilibrium state, such that over-limiting currents cannot be observed. However, it leaves a clear and unique EOI signature: EOI can be observed in both depletion layers and in $\tilde{i}(t)$ [**Figure 5**(b)]. The response of non-ideal bipolar systems is comprised of several characteristics of the unipolar and ideal bipolar responses. This includes the formation of EOI on both sides, but the eventual decay of EOI on one side and the appearance of over-limiting currents [**Figure 2**]. In this Part, we have elucidated the time transient response of unipolar and bipolar systems subject to over-critical voltages, which, in the right conditions, lead to over-limiting currents. Nonetheless, whether EOI is desired or not and whether it can be leveraged is subjective based on the application (desalination, energy harvest, biosensing, and more). Optimization of the effects of EOI and delineating its variations on the performance of each application require a lot more fine-tuning associated with the particular application. Yet, this work focuses on more objective and robust issues. Namely, we show, based on the system parameters, whether or not EOI appears in bipolar systems – without this knowledge, an advanced and accurate experimental protocol cannot be formulated.

It is our hope that the systematic approach undertaken in this work will provide much-needed novel insights into the behavior of these systems as well as how to (optimally) characterize them through their very unique time-dependent signatures that can be observed or measured. These time-dependent signatures, in conjunction with fluorescent microscopy, can be used to confirm or validate whether or not the EOI forms in experimental bipolar nanofluidics systems.

**Acknowledgments.** We thank Oren Lavi for several in-depth discussions and very useful comments on how to streamline the results of this two-part work.

**Competing interests.** The authors declare no conflict of interest.

**Funding statement.** This work was supported by Israel Science Foundation grants 337/20 and 1953/20. We acknowledge the support of the Ilse Katz Institute for Nanoscale Science & Technology.

**Data Availability Statement.** Raw data will be made available upon request from the corresponding author (Y.G.).

**Supplementary Material and Movies.** Supplementary material documents and supplementary movies intended for publication have been provided with the submission. These are available at https://doi.....

**Author ORCIDs.**

Ramadan Abu-Rjal https://orcid.org/0000-0002-1534-9710
Yoav Green https://orcid.org/0000-0002-0809-6575